# To Improve Cyber Resilience, Measure It

Alexander Kott, US Army CCDC Army Research Laboratory, USA

Igor Linkov, US Army Engineer Research and Development Center, USA

Keywords: performance measures, cyber resilience, cyber survivability, adversary emulation, response and recovery

We are not very good at measuring – rigorously and quantitatively – the cyber security of systems. Our ability to measure cyber resilience is even worse. And without measuring cyber resilience, we can neither improve it nor trust its efficacy.

Even the meaning of the term cyber resilience is often misunderstood. Cyber resilience refers to the ability of systems to resist and recover from, or adapt to, a cyber compromise (DoD, 2014). A few attempts to measure cyber resilience have focused on the ability of systems to resist well-defined and predictable threats, that is, avoid compromise, which is part of the traditional risk assessment and management process (Linkov and Trump, 2019). What is measured in this case is the probability of the failure of system components in response to different threat scenarios with the goal of hardening the system so the compromise does not happen. This is not exactly resilience. The key notion of cyber resilience is acceptance of cyber compromise as a likely event, and the system suffering as a result; the focus is on the system's ability to recover and adapt, and not just resist. Cyber resilience characterizes what happens *after* an adverse event, and requires preparedness for both known and unknown threats (Kott and Linkov 2019).

A number of mechanisms arguably supporting system resilience have been proposed and already applied in real-world systems, and more are being developed. Unfortunately, most of these mechanisms just harden the system to prevent compromise. One might presume that the more mechanisms, the more secure our systems will be. This has not been true for cybersecurity, nor do we believe it is the case for cyber resilience. In fact, each additional mechanism carries with it the risk of expanding the surface of a successful attack. It also introduces the risk of interfering with the work of other mechanisms already built into the system, reducing their effectiveness.

There is also a risk that a response action taken by a security-support mechanism will lead to cascading failures and ultimately "break" the computer, disabling critical software or corrupting important data. As such, control mechanisms designed to support security can actually be detrimental to the goal of resilience. It is difficult to know if we are improving or degrading cyber resilience when we add another control, or a mix of controls, to harden the system. The only way to know is to specifically measure cyber resilience with and without a particular set of controls. What needs to be measured are temporal patterns of recovery and adaptation, and not time-independent failure probabilities. The question we address is how to measure cyber resilience.

**Cyber resilience and its Importance**

The ability to restore a system's functionality after a cyber-compromise is important for a broad range of systems. Cyberattacks on critical infrastructure – including water supplies, energy and communication networks, and health care facilities – are highly consequential (Linkov et al., 2013). A major health services organization, subjected to a ransomware attack, was reduced to manual methods of data handling, over a significant period of time (Landi, 2020). Such attacks can produce massive damage to the economic well-being of an organization and to broader society, and even endanger human lives. Ironically, impressive examples of cyber resilience may come from malware operations. For instance, the notorious TrickBot – a botnet that conducts ransomware attacks – has demonstrated an agile and effective recovery after competent malware-fighting organizations attempted to dismantle the botnet (Waldman, 2020).

When the adverse cyber event occurs, the system absorbs the threat and its functionality begins to degrade; various mechanisms and processes (which may or may not include human actions) engage in absorbing the negative impacts, and then in recovering the system's functionality (Figure 1). Cybersecurity is focused on hardening the system to prevent such degradation; the focus is on the degree of degradation.

In contrast, cyber resilience is focused on recovery. The recovery may be complete or partial, and in some cases may involve adaptations that would improve functionality or resilience to future adverse events. Naturally, cyber resilience depends on aspects of the system, such as its design, controls, preparation, anticipation, training, etc., that occur *before* the adverse event. However, the ultimate assessment of cyber resilience starts with acknowledging the inevitability of adverse events: when the system is affected, functionality is degraded, and the focus is on the speed of recovery. As illustrated in Figure 1, a more resilient system would exhibit a greater area under the curve (AUC) – the integral of retained system functionality $F(t)$ over the total mission time $T_m$. If so, we might try to define the resilience quantity as $R = (AUC)/(F(0)*T_m)$, but we will return to this point later.

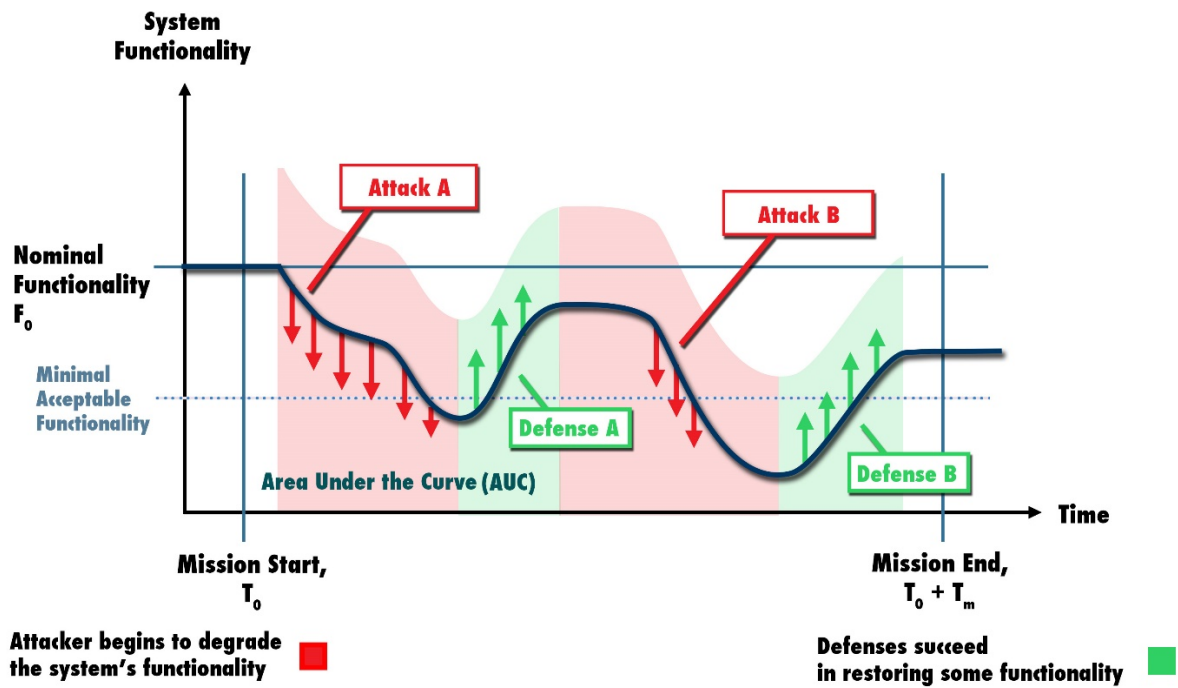

Figure 1. Stages of cyber resilience

When (not if) a cyber incident occurs, specialized organizations such as security operation centers, managed security service providers, incident response providers, become the agents of resilience. They identify the nature of the compromise, isolate and contain the compromise, engage redundant computing resources, sanitize the affected equipment, reinstall the software, and restore data from backups. All these steps require substantial – and expensive – human expertise. Worse yet, these processes can take precious time – often hours or even weeks.

Other situations demand much faster responses, and resilience that relies on human responders may be untenable for some categories of use cases. Criminals or irresponsible pranksters are able to take control of cars traveling at high speed, or planes in the air, which may constitute a mortal threat to the vehicle's passengers and others interacting with those systems (Ring, 2015). In such cases, waiting for a human incident response team will not do. Instead, such systems need an onboard intelligent autonomous agent capable of taking the necessary response and recovery actions, with response times on the order of seconds or even less (Kott & Theron, 2020).

The need for super-humanly fast resilience is even more pressing in warfare, especially the highly computerized warfare likely to characterize the near future. To maintain a meaningful degree of effectiveness, a missile defense system may have only a few seconds to react to a cyber-compromise and recover. Or, consider tactical ground mobile assets, in future active ground combat. These could include tanks, combat robots (both ground and air vehicles), intelligent sensors, intelligent munitions, and so on. These assets will tend to operate in relatively close proximity to enemy forces, which in turn will entail relative ease of both physical and electromagnetic access, with some probability of physical capture of assets and human soldiers by the adversary. In other words, the probability of cyber compromise – serious enough to degrade the system's functionality below an acceptable level -- is relatively high.  If such an incident occurs, whose responsibility will it be to address it? Soldiers on manned assets are too busy with other life-and-death tasks and possess neither the time nor the skills to deal with a cyber-compromise. And unmanned assets by definition don't have on-board human defenders.

Another approach could be a cyber operations center, with well-trained cyber warriors, who would address such compromises ~~remotely~~. However, this is unlikely to be successful: the future battlefield will see highly contested networking, intermittent connectivity, and the need to minimize radio emissions. Thus, we cannot rely on remote recovery. The remaining solution is that resilience must depend on artificially intelligent agents residing onboard on-the-battlefield assets, and provide real-time response and recovery during the mission (Kott et al., 2019). Nevertheless, the acceptance of even recommended decision by such systems, not to mention allowing automated response to major compromises, requires the ability to provide a degree of confidence associated with AI-driven interventions to be actionable (as argued in Linkov et al., 2020), which is yet again require a good measures and metrics.

It's not that we as a community don't work hard to improve cyber resilience – we do. Let's consider some examples of attempts to enhance cyber resilience. The U.S. National Institute of Standards and Technology (NIST) published a comprehensive catalog of potential techniques to improve the cyber resilience of systems (Ross et al., 2018). Some of these techniques already contribute to the cyber resilience of commercial products. For instance, micro-segmentation is a widely-used approach which improves resilience by slowing down cyber intruders as they attempt to navigate through the system. Cyber deception is an active topic of academic research (Rowe and Rrushi, 2016). RHIMES is a research program funded by the Office of Naval Research (ONR) which uses a range of detection and recovery techniques for securing cyber-physical systems from cyber attacks (Pomerleau, 2015). A NATO research group has proposed a reference architecture for an Autonomous Intelligent Cyber-defense Agent (AICA) that would reside on a system, continually assessing adversarial activity on the system, and autonomously plan and execute mitigation and recovery actions (Kott et al., 2019).

**Better resilience calls for measurements**

Here by measuring we understand (consistent with prevailing literature on the topic of measurements) quantifying, objectively and empirically, a specified attribute of an actual phenomenon experienced by an actual (or representative) system of interest. Measurements must be as empirical and "physical" as possible, even in a cyber-world. Certain methods and measures are often confused for measurements. For instance, great progress has been made in the modeling and simulation of cyber phenomena. But these are not measurements. We do qualitative assessments, and we have check lists and metrics for these assessments – very appropriate and necessary for decision-making, but require underlying measurements. We use Red Teams – very important, but these, again, are not measurements.

In seeking potential approaches to measuring cyber resilience, it may be helpful to consider a well-developed measurement process in another field, physics: measuring the properties of materials. These measurements (tests) often involve the application of a destructive effort (e.g., a cyclic load of a given magnitude) applied to a sample of material that performs an abstracted form of a useful duty (e.g., withstanding a large number of bending cycles in case of measuring the fatigue resistance property). These tests may also involve varying the load and quantifying the extent (e.g., the number of cycles tolerated before the destruction occurs) to which the sample is able to accomplish its duty under the load.

Examining this process in the realm of materials science can provide insights into possible features of measurement techniques for cyber resilience. First, one should define an abstracted yet representative mission of the system and critical functions necessary to sustain the mission being measured. Second, devise a way to apply a cyber adversity of a given type (both as a targeted and random attack), and to vary the magnitude (clearly, research is also needed on how to measure the magnitude of the cyber pressure). Third, develop the means by which the system executes its prescribed mission while experiencing the cyber load. Fourth, quantify the extent to which the system is able to perform its mission. For instance, functionality averaged over the time of the mission (see Figure 1) might constitute a way to quantify resilience. However, research is needed to quantify "functionality" and to understand how to account for the time-variable cyber load and system response. Finally, provide tools and processes that ensure such experiments can be repeated consistently and objectively, in conjunction with simultaneous modeling and simulation, towards obtaining sufficient evidence-based data for confident decision making (see next section).

For example, having measured the fatigue resistance, engineers can use this measurement to predict if and when a particular part will fail. In a somewhat similar way – although such analogies are never perfect – cyber resilience measurements help the designers or operators to estimate the suitability of a system (along with its resilience mechanisms) for a given mission.

**Confidence in Measurements and Metrics**

Unlike the precise measurements provided in physical science, measuring cyber resilience is in its infancy. Decisions on converting knowledge and intuition regarding the recovery and adaptation of cyber systems in response to threats into management decisions and policy will rely on a growing volume of increasingly diverse measurements. These measurements can be collected by different modes, compounded with experiments or models, and can point in different directions regarding the system functionality. How can cyber professionals make confident decisions on appropriate courses of action given the connection of multiple individual measures associated with threat absorption, recovery, and adaptation of the system in response to cyber threats?

One approach may be to develop a set of criteria that would ensure decision-maker confidence in the reliability of the methodology used in obtaining a meaningful measurement. For example, the Bradford-Hill (BH) criteria (Hill, 1965), originally developed for the evaluation of causality of associations observed in epidemiological studies, and more recently evolved to increase consistency for other determinations (e.g., Becker et al., 2015), provide a useful approach for evaluating resilience metrics. The nine "aspects of association" that Hill discussed include strength of association, consistency, specificity, temporality, biological gradient, plausibility, coherence, experiment, and analogy. More work is required to develop criteria relevant to cyber systems, but the following may be a starting point:

- Repeatability: we anticipate that, for a given system and measurement tools, a repeated series of measurements should yield approximately the same value of resilience.
- Consistency with respect to missions: we anticipate that similar (but not identical) missions should yield reasonably similar values of the resistance quantity $R$. There might be some important reasons for a difference (e.g., there is a subtle but major difference between the two missions), but in the absence of a plausible explanation we must question the methodology.
- Monotonicity with respect to defenses: we anticipate that significantly stronger on-board cyber-defenses should yield a higher value of $R$. This presumption might be violated (e.g., the putative better defense may in fact introduce new weaknesses), leading to further investigation of the causes.
- Monotonicity with respect to attacks: we anticipate that significantly stronger cyber-attacks should yield a lower value of $R$. Again, this presumption might be violated for a legitimate reason (e.g., the putative stronger attacks may turn out to be ineffective against the system's defenses), and again we should seek a plausible explanation or, in its absence, improve our methodology.

The second approach may be to integrate large volumes of information and multiple measurements to express confidence levels quantitatively. In statistics, the confidence level associated with data is well defined, but in emerging fields confidence must integrate data and technical judgments and value calls by decision makers and commanders. Challenges – and the need for further research – related to confidence in cyber resilience will grow as we see greater use of artificial intelligence and autonomous agents, integrated in various ways with human hierarchical guidance or in-the-loop decision. The interpretation of the performance of individual measures with respect to achieving specific cyber missions as well as translating inevitable uncertainty and variability associated with cyber resilience metrics can be facilitated by the use of weight of evidence (WOE) evaluation. WOE is an approach that, by means of qualitative or quantitative methods, integrates individual lines of evidence (i.e., metrics in this case) to form a conclusion (Linkov et al., 2015) and has been widely used in the field of risk assessment to collate heterogeneous information and justify the selection of regulatory benchmarks and appropriate courses of action. Recent developments in the field of WOE (2015) are focused on quantitative (e.g., Bayesian) and semi-quantitative (e.g. Multi Criteria Decision Analysis) methods as a foundation for information fusion.

**The history of technology tells us so**

We are not offering a solution to this problem, at least not yet. Instead, we are inviting the community of cyber researchers and practitioners to engage in conversation and exploration. Clearly, we cannot reliably improve what we cannot measure. Sciences and engineering fields blossomed only when measurement tools appeared. No technical discipline has achieved a degree of maturity without developing techniques, tools, and processes for objectively, rigorously, and quantitatively measuring the attributes of phenomena occurring in the systems of that discipline. Cyber resilience is no exception.

**Disclaimer**

**About the authors:**

*Vita: Alexander Kott is the Chief Scientist of the US Combat Capabilities Development Command's Army Research Laboratory. Contact him at alexander.kott1.civ@mail.mil.*

*Vita: Igor Linkov leads the Risk and Decision Science Team at the US Army Engineer Research and Development Center, US Army Corps of Engineers. Contact him at Igor.Linkov@usace.army.mil.*